\PassOptionsToPackage{warn}{textcomp} 

\documentclass[acmsmall,%
  nonacm, usenames, dvipsnames]{acmart}
\renewcommand\footnotetextcopyrightpermission[1]{} 
\usepackage{import}

\usepackage{amsmath}
\usepackage{graphicx}
\usepackage{float}
\usepackage[document]{ragged2e}
\usepackage{wrapfig}
\usepackage{stackrel}
\usepackage{extpfeil}
\usepackage{tikz}
\usepackage{tikz-cd}
\usepackage{tikz-qtree}
\usetikzlibrary{matrix}
\usetikzlibrary{automata, positioning, arrows,shapes.geometric}
\usetikzlibrary{babel}
\usepackage{turnstile}
\usepackage[utf8]{inputenc}
\usepackage[T1]{fontenc}
\usepackage{graphicx}
\usepackage{grffile}
\usepackage{longtable}
\usepackage{wrapfig}
\usepackage{rotating}
\usepackage[normalem]{ulem}
\usepackage{textcomp}
\usepackage{capt-of}
\usepackage{listings}
\usepackage{subfigure}
\usepackage{bm}
\usepackage{comment}
\usepackage{mathtools}
\usepackage{hyperref}

\newcommand{\xtransitsto}[2]{{\underset{#2}{\xrightarrow{#1}}}}

\newcommand{\xto}[2]{{\xtransitsto{#1}{#2}}}

\def\to{\rightarrow}

\def\too{\longrightarrow}

\def\ebox\Box
\def\emp{\emptyset}

\newcommand{\benum}{\begin{enumerate}}
\newcommand{\eenum}{\end{enumerate}}
\newcommand{\bdes}{\begin{description}}
\newcommand{\edes}{\end{description}}

\newcommand{\bt}{\begin{theorem}}
\newcommand{\et}{\end{theorem}}
\newcommand{\bl}{\begin{lemma}}
\newcommand{\el}{\end{lemma}}
\newcommand{\bp}{\begin{prop}}
\newcommand{\bd}{\begin{defn}}
\newcommand{\ed}{\end{defn}}
\newcommand{\brem}{\begin{remark}}
\newcommand{\erem}{\end{remark}}
\newcommand{\bxr}{\begin{exercise}}
\newcommand{\exr}{\end{exercise}}
\newcommand{\bxm}{\begin{example}}
\newcommand{\exm}{\end{example}}

\newcommand{\beqa}{\begin{eqnarray*}}
\newcommand{\eeqa}{\end{eqnarray*}}
\newcommand{\bc}{\begin{center}}
\newcommand{\ec}{\end{center}}
\newcommand{\bcent}{\begin{center}}
\newcommand{\ecent}{\end{center}}

\newcommand{\bcor}{\begin{corollary}}
\newcommand{\ecor}{\end{corollary}}
\newcommand{\bds}{\begin{defns}}
\newcommand{\eds}{\end{defns}}
\newcommand{\brems}{\begin{remarks}}
\newcommand{\erems}{\end{remarks}}
\newcommand{\bxrs}{\begin{exercises}}
\newcommand{\exrs}{\end{exercises}}
\newcommand{\bxms}{\begin{examples}}
\newcommand{\exms}{\end{examples}}
\newcommand{\bfig}{\begin{figure}}
\newcommand{\efig}{\end{figure}}

\newcommand{\id}[1]{\mathit{#1}}

\newcommand{\da}[1]{\bigg\downarrow\raise.5ex\rlap{\scriptstyle#1}}
\newcommand{\ua}[1]{\bigg\uparrow\raise.5ex\rlap{\scriptstyle#1}}
\DeclareMathSymbol{\shortminus}{\mathbin}{AMSa}{"39}
\DeclareMathSymbol{\sm}{\mathbin}{AMSa}{"39}

\usepackage{bm}

\makeatletter
\DeclareRobustCommand*\cal{\@fontswitch\relax\mathcal}
\makeatother

\tikzset{
node distance=3cm, 
initial text={},
every edge/.style={ %
draw,
->,         
>=stealth', 
auto,
semithick,
bend angle=15
}
}

\renewcommand\paragraph[1]{}

\mathchardef\mhyphen="2D
\AtBeginDocument{%
  \providecommand\BibTeX{{%
    \normalfont B\kern-0.5em{\scshape i\kern-0.25em b}\kern-0.8em\TeX}}}

\begin{document}


\title[Algodynamics]{Algodynamics: Interactive Transition
  System approach to Step-wise Refinement in the Design of
  Algorithms}



\title[Algodynamics]{Algodynamics: Teaching Algorithms using Interactive Transition Systems}

\author{Venkatesh Choppella}
\affiliation{%
  \institution{IIIT Hyderabad}
  \country{India}
}

\author{Viswanath Kasturi}
\affiliation{%
  \institution{IIIT Hyderabad}
  \country{India}
}

\author{Mrityunjay Kumar}
\affiliation{%
  \institution{IIIT Hyderabad}
  \country{India}
}

\author{Ojas Mohril}
\affiliation{%
  \institution{IIIT Hyderabad}
  \country{India}
}








\renewcommand{\shortauthors}{Choppella, Kasturi, Kumar, Mohril}

\begin{abstract}
The importance of algorithms and data structures in computer
science curricula has been amply recognized.  For many
students, however, gaining a good understanding of
algorithms remains a challenge.

Because of the automated nature of sequential algorithms
there is an inherent tension in directly applying the
`learning by doing' approach.  This partly explains the
limitations of efforts like algorithm animation and code
tracing.

Algodynamics, the approach we propose and advocate, situates
algorithms within the framework of transition systems and
their dynamics and offers an attractive approach for
teaching algorithms.  Algodynamics starts with the premise
that the key ideas underlying an algorithm can be identified
and packaged into interactive transition systems.  The
algorithm when `opened up', reveals a transition system,
shorn of most control aspects, enriched instead with
interaction.  The design of an algorithm can be carried out by
constructing a series of interactive systems, progressively
trading interactivity with automation.  These transition
systems constitute a family of notional machines.

We illustrate the algodynamics approach by considering
Bubblesort.  A sequence of five interactive transition
systems culminate in the classic Bubblesort algorithm.  The
exercise of constructing the individual systems also pays
off when coding Bubblesort: a highly modular implementation
whose primitives are borrowed from the transition systems.
The transition systems used for Bubblesort have been
implemented as interactive experiments.  These web based
implementations are easy to build.  The simplicity and
flexibility afforded by the algodynamics framework makes it
an attractive option to teach algorithms in an interactive
way.

\end{abstract}


\maketitle
\fancyfoot[R]{October, 2020}

\section{Introduction}
\label{sec:intro}

\paragraph{Algorithms in the ACM curriculum}

Algorithms form an essential component of recommended
curricula in computer science\cite{cs2013}, computer
engineering\cite{10.1145/792548.611915} and software
engineering\cite{10.1145/2965631}.  On the other hand,
students continue to face difficulty with algorithmic
problem solving: in problem formulation, notation for
expressing solutions, logical reasoning, tracing the
execution, and understanding the behaviour of
their solutions\cite{medeiros_systematic_2019}.

Investigation into the sources of this difficulty has mostly
focused around programming
\cite{winslow_programming_1996,robins_learning_2003,medeiros_systematic_2019}
rather than algorithms directly.  This is not surprising;
algorithms are ultimately implemented as programs.  Mental
models are useful in solving problems
\cite{halasz1983mental}, but in the absence of a clear,
structured and simple notation and a standard abstract
model, students try to construct ad-hoc mental models built
from programming constructs or pseudo-code
\cite{10.1207/s15327051hci0902_3}.

%
The central role of an abstract machine in defining the
concept of an algorithm has been emphasized by
Denning\cite{10.1145/2998438}.  Other recent efforts have
also sought to reconsider the role of notional machines,
which abstract the workings of a particular programming
language, paradigm or hardware
architecture\cite{du_boulay_difficulties_1986,
  10.1145/2483710.2483713, guzdial2019notional}. %

%
%
We take the position that algorithms need their own,
tailor-made abstract machines.  An algorithm-specific
abstract machine, as opposed to a `general computer',
directly works with the abstractions inherent in the problem
and the high-level operations used by the algorithm.

It is widely understood that interaction is an essential
component of `active learning' and `learning by
doing.'\cite{10.1145/973801.973818, blasco2013using}.  The
student's own experience with computation is mostly defined
by interaction with digital devices like mobile phones.  It
is natural, therefore to expect abstract machines, as well,
to incorporate interactivity.  

Computer scientists already have a powerful formalism to
study interactive abstract machines: transition systems.
Such systems exhibit deterministic and non-deterministic
behaviour, are terminating or non-terminating, and can model
both sequential and concurrent systems.  In computer
science, they are used in the study of verification and
control of reactive and embedded systems
\cite{tabuada-book-2009,lee-seshia-book-2e-2015}. Their
potential as abstract machines for building mental models of
algorithms and data structures execution remains mostly
unexplored.

\begin{figure}
    \includegraphics[width=3.0in]{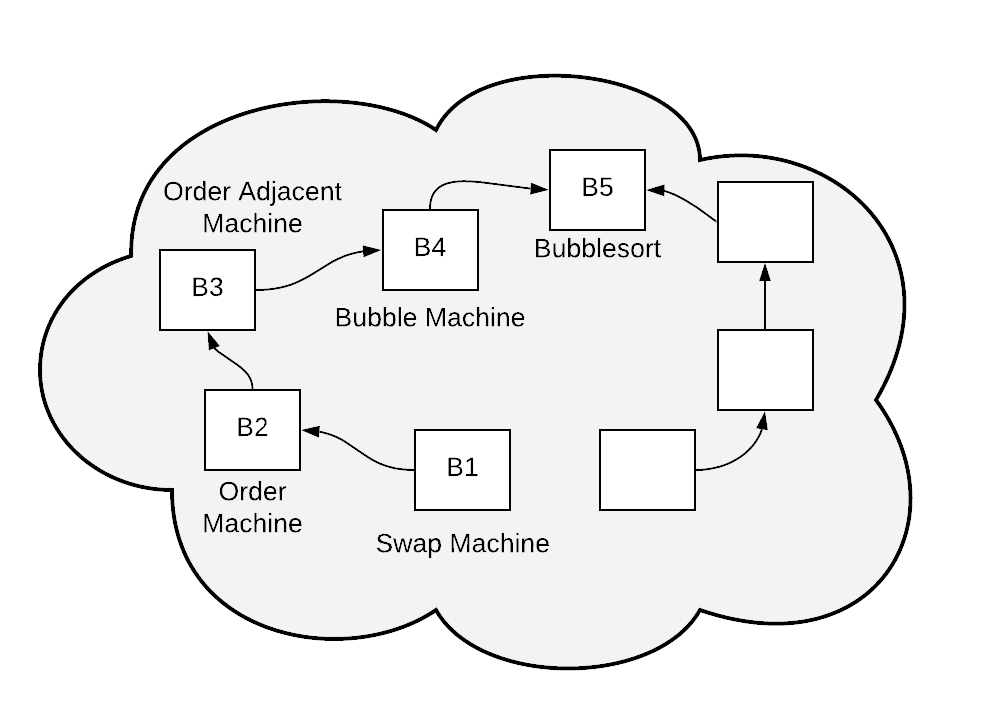}
    \caption{Approaching an algorithm (Bubblesort) via a
      sequence of transition systems (machines). The diagram
      shows two of the many possible design
      pathways. \label{fig:algodynamics}}
  \end{figure}

In this paper, we propose {\em Algodynamics}, a pedagogical
framework that situates algorithms in the design space of
interactive transition systems.  In trying to bring
interaction to algorithms, however, we face the following
apparent challenge: an algorithm is non-interactive, (i.e.,
automated), sequential and terminating.  Its notional
machine inherently does not lend itself to interaction.
However, seen as an transition system, an algorithm consists
of a trivial action: {\sf next} and is situated at the
non-interactive end of the spectrum of transition systems.
In algodynamics, the understanding of an algorithm is
approached via a sequence of interactive transition systems
culminating in the algorithm.  Figure~\ref{fig:algodynamics} tries to capture this idea. %

By bringing in transition systems, we bring in an alternative
notational and reasoning framework that is abstract, precise
and compact.  The framework allows the expression of mental
models that align closer to the problem and the high-level
operations rather than the implicit architectural model of a
programming language (functions, while loops, etc.)

Students trace their programs to understand how their
algorithm runs.  Algodynamics connects it to a sequence of
transitions.  Because the elements of the trace operate on
higher abstract actions, the cognitive load on the student
is reduced.

At each stage in the process leading to the construction of
the algorithm, we have a complete, interactive system that
the student can play with, reason about and develop insight
about the problem and the solution and explore other
solutions and problems in the vicinity.  This approach
encourages creative experimentation (the highest level of
Bloom's taxonomy).  In the programming part of the lab, the
student is simply coding, in a systematic manner, the
transition systems she has already understood.

In the rest of the paper, we review the literature on the
difficulties of teaching and learning algorithms through
programming(Section~\ref{sec:related-work}) and then include
a short self-contained, introduction to transition systems
and algodynamics (Section~\ref{sec:ts}).  We illustrate the
algodynamics approach by exploring the Bubblesort algorithm
(Section~\ref{sec:bubblesort}).  Conclusions and future work
are presented in Section~\ref{sec:conc}.

\section{Related Work}
\label{sec:related-work}

In this section, we briefly review the challenges students
face that are related to problem solving, programming and
algorithms.

Learning programming continues to be difficult, as evidenced
by various studies on pass rates of introductory programming
courses across various countries \cite{luxton_pass_2019,
  watson_failure_2014, bennedsen2007failure}.  The
difficulties of novice programmers has been widely
researched \cite{du_boulay_difficulties_1986,
  spohrer_novice_1986}, including in a recent literature
survey by Medeiros\cite{medeiros_systematic_2019}, who
groups the difficulties of novice programmers into 4
categories: 1) problem formulation, 2) solution expression,
3) solution execution and evaluation, and 4) behavior.

Difficulties in problem formulation include problem solving,
the abstract nature of programming, and algorithmic and
logical reasoning. These resonate with Winslow's discussion
on novice vs. expert \cite{winslow_programming_1996}:
\textit{"Experts tend to approach a program through its
  objects rather than control structures... experts use
  algorithms, rather than a specific syntax, abstracting
  from a particular language to the general concept."}
Robins\cite{robins_learning_2003} lays out the novice
vs. expert differences in great detail, citing several
studies.  He indicates \textit{"the most important deficits
  relate to the underlying issues of problem solving,
  design, and expressing a solution/design as an actual
  program."} Fix \cite{fix_mental_1993} studies mental
representation of programs by novices and experts and finds
that the novice's representation lacked the characteristics
evidenced by the expert's representation.

Program comprehension is another area cited as difficult for
novices, as evidenced by their ability to trace the program
and also explain it in English
\cite{lister_multi-national_nodate}. Vainio
\cite{vainio_factors_nodate} studies the causes for poor
tracing skills and describes 4 specific difficulties
students face, including \textit{Inability to use external
  representations} and \textit{inability to raise
  abstraction level}.  Lopez \cite{lopez_relationships_2008}
and Lister \cite{lister_further_nodate} also conclude that
those who can't trace well also can't explain the code, and
those who perform reasonably well at code writing have also
acquired the ability to trace and explain.

Summarizing these observations, it seems clear that students
need to acquire the following skills to become better at
problem solving: 1) \textbf{Abstraction and Modeling} -
ability to model the problem and think at an abstract level,
using external representations as needed, 2) \textbf{Program
  comprehension and tracing} - ability to understand the
program by tracing and explaining in English, and 3)
\textbf{Algorithmic reasoning} - ability to reason logically
rather than via program syntax and control structure.


An algorithm course in an undergraduate program is expected
to address the above concerns. However, given the challenges
above, it is important to first review the ways in which
the teaching of algorithms is approached and to then propose
pedagogical improvements that address the challenges above.

The concerns with the way algorithms are taught are not
new. Levitin \cite{levitin2000design} argued to reconsider
the way we teach design and analysis of algorithms and
incorporate more problem-solving, almost 20 years ago.
While the constructivist approach has been espoused for
teaching programming \cite{ben2001constructivism,
  10.1145/1095714.1095771} and constructionism
\cite{harel1991constructionism} for experiential learning,
we do not see evidence in the literature of knowledge
construction approaches when teaching algorithms. Leading
textbooks on algorithms \cite{cormen2009introduction,
  sahni1978fundamentals, skiena1998algorithm,
  sedgewick1988algorithms} teach algorithms in a fait
accompli manner - algorithms are presented in real code or
pseudo-code, and the applicable strategy is pointed out
('divide and conquer', 'backtracking', 'greedy', 'dynamic',
etc.).  Computer Science Curricula 2013 \cite{cs2013}
approaches Algorithms and Complexity Body of Knowledge along
similar lines. Use of code makes algorithm understanding
hard for the same reason novice programmers find programming
hard. Use of pseudo-code makes it only somewhat easier due
to the lack of notational and semantic standards
\cite{cutts2014code}. Such an approach to teaching doesn't
help the students in constructing their own knowledge about
algorithm design.

A common and frequently discussed, approach in the
literature for algorithm teaching is algorithm visualization
(AV) \cite{shaffer_algorithm_2010, alecha_learning_2009,
  alharbi_integrated_2010, vrachnos_design_2014,
  amershi_pedagogy_2008}. While AV does offer better
engagement and explainability, the outcomes and usage
haven't been significant enough to label it as an effective
approach to teaching algorithms
\cite{urquiza-fuentes_survey_2009, vegh_algorithm_2017,
  hundhausen_meta-study_2002}.  One of the challenges could
be the limited role of interactivity in these
visualizations. While there is evidence that adding
interactivity to visualizations can positively impact
understanding, \cite{wang_impact_2011,
  patwardhan_when_2015}, most of the algorithm
visualizations available today lack interactivity.

An additional impediment that makes it hard harness
interactivity and create game-like environments for
algorithm teaching is that algorithms are inherently
automatic and non-interactive.
 
We believe that an effective approach to teaching algorithms
needs to focus on answering these questions: 1) How do we
enable students to create abstract mental representations of
algorithms? 2) How do we help students trace better and
reason about the program? and 3) How do we let the students
interact with the algorithm infrastructure and learn the
principles of algorithm design through such interactions?

In his 1971 seminal paper \cite{wirth2001program}, Niklaus
Wirth says: \textit{"Clearly, programming courses should
  teach methods of design and construction, and the selected
  examples should be such that a gradual development can be
  nicely demonstrated.}

In this paper, we take inspiration from the approach Wirth
suggests for programming and apply it towards the design and
construction of algorithms.  The method we employ is interactive
transition systems.  The gradual development is demonstrated
by the sequence of transition systems, each addressing a
design concern and deciding on a data representation or
control strategy.  This way, students understand the nature
of the algorithm and also engage with it sufficiently to
develop their own understanding.

\section{Transition Systems and the Algodynamics Approach}
\label{sec:ts}
A {\em transition system} is a model to understand how
quantities of interest change when subjected to outside
forces called inputs or actions.  The set of all possible
values of the tuple of quantities constitute the state space
of the system. {\em Dynamics} studies how the state changes
when subjected to a series of actions over time.
\emph{Algodynamics} models algorithms as transition systems
and studies their dynamics.  In the algodynamics approach,
finding a solution to a computation problem reduces to
finding a sequence of the actions that steer a system from
an initial state to its final state, the desired state.

Algodynamics explores the space between interactive systems
and algorithms.  In this space exist several transition
systems, each presenting itself as an interactive laboratory
that reveals some important aspect of the algorithm.  The
laboratory keeps the student engaged and encouraged to
discover intereactively new ways to solve a problem. Because
a transition system is interactive, the student can
\emph{play} with the system by choosing an action (akin to
making a move in a game) and seeing its effect on the
system's state.  What sequence of actions to apply is left
to the student, and the student may devise her own
\emph{strategy} to choose these moves.  From merely
`observing', or `tracing' how an algorithm runs, the student
now takes control and steers the system towards the
solution.  The student is now `learning by doing.'

The opposite of `interactive' is \emph{`automated'}. A
transition system is said to be automated if there is only
one action that can be performed on the state.  Algorithms
are automated systems.  In the algodynamics approach, the
student is introduced to a gradually developed series of
increasingly less interactive transition systems which
culminate in the algorithm.  At each stage the student is
trading interaction for automation.  This allows the student
to see the design of an algorithm emerge in an incremental
manner.

\subsection{Brief introduction to Transition Systems}
Our definition of a transition system is one that has now
become
standard\cite{2008-book-baier-katoen,tabuada-book-2009,2017-book-belta-et-al}.
A transition system (TS) is a six-tuple $(X,X_0,,U,\delta,
Y,h)$ where $X$ is a set called the state space; $X_0$ is a
subset of $X$ called the set of initial states; $U$ is a set
called the set of actions; $\delta: X \times U \to 2^X$ is
the transition function or {\em dynamics}; if $\delta(x,u) =
x'$ we write $x \stackrel{u}{\too}x'$, which is called a
transition.  $Y$ is a set called the view (or observation)
space and $h: X \to Y$ is a map called the view map.  A TS
is {\em automated} if the action set is a singleton.  The
set of states, observation and actions could each be finite,
infinite or continuous.  A system is said to be {\em
  deterministic} if $|\delta(x,u)| \leq 1$ for all $x$ and
$u$.  For a deterministic system, whenever $x
\stackrel{u}{\too}x'$, $x'$ is unique.  A state $x \in X$ is
said to be terminating if $\delta(x,u) = \emptyset$ for all
$u \in U$.  A run is a sequence of linked transitions $x_0
\stackrel{u_1}{\too} x_1 \stackrel{u_2}{\too} x_2 \cdots
x_{n-1} \stackrel{u_n}{\too} x_n.$ For a run as above the
sequence of states $(x_0,x_1,\cdots,x_n)$ is called a
trajectory and the sequence of observations in the set $Y$,
namely, $(h(x_0),h(x_1),\cdots,h(x_n))$ ~~~~ is called a
trace.  A TS is {\em terminating} if it has no infinitely
long runs.  

A TS is generally interactive. A user can choose an initial
state $x_0$, and a sequence of actions $u_1,u_2,\cdots,u_n$
to construct a run and see what the trajectory looks
like. She may like to choose actions that result in a
specified trajectory. As remarked earlier a TS is said to be
automated if there is only one action in $U$. In such a case
the unique action is generally denoted by $next$ and there
is no more any choice for the user in the construction of
trajectories. The choice of an initial state determines the
entire trajectory.

A computation problem is specified by choosing a view space
$Y$, an initial subset $Y_0$ of $Y$, a final subset
$Y_{\omega}$ of $Y$, and a map $f: X_0 \to Y_{\omega}$
called the specification map.  An interactive solution of a
computation problem is a TS whose traces begin with some
$y_0 \in Y_0$ and end in $f(y_0) \in Y_{\omega}$.  If an
interactive solution can be designed to be automated,
terminating and deterministic, then the solution is called
an {\em algorithm}.


In Algodynamics, we take the specification of a computation
problem as stated above and try to construct tentatively an
interactive solution with some wide and suitable choice of
data types and actions to construct $X$ and $U$, followed by
the full transition system.  Then we progressively refine it
to generate a sequence of interactive solutions that ends
with an algorithm.

\section{Illustrating the Algodynamics approach with Bubblesort}
\label{sec:bubblesort}


\paragraph{Algodynamics approach}

We apply the algodynamics approach to explore the
specification of the sorting problem and the Bubblesort
algorithm.  Our concern here is not Bubblesort's efficiency
(which is poor), nor its persistent use in algorithm
pedagogy despite its many
shortcomings\cite{10.1145/792548.611918}, but its mechanics,
which is simple, but not trivial.

The sorting problem may be specified as building an
algorithmic system with initial observation $a^0$, an
$n$-sized array of numbers to be sorted, and in whose
terminal state, we observe $\id{sort}(a^0)$, its sorted
permutation.  The goal is to build an algorithmic transition
system for Bubblesort and implement it as a program.

\subsection{A sequence of transition systems for Bubblesort}
\label{sec:seq-of-ts}

We suggest a sequence of five transition systems $B_1$ to
$B_5$ to explore the design of Bubblesort.  The sequence is
but one of many possible trajectories in the design space
that contains Bubblesort.  Each system in the sequence $B_1$
to $B_5$ highlights a key decision in the design of
Bubblesort.  We present sample runs, and briefly note the
key properties of the system.  We include screenshots of
online experiments for some of the systems.  A full
exposition should include mathematical arguments of
correctness.  We elide them here in the interest of brevity.

{\bf State variables and Notation}: We use $a$ to denote the
array state variable.  The size of $a$ is assumed to be $n$
and element indexing is zero-based.  In addition, we assume
array index variables $0 \leq i < j < n$ and $0 \leq b \leq
n$ which appear as part of the state vector in specific
transition systems.

\subsubsection{Transition System $B_1$: ``Swap Machine''}
\label{subsubsec:b-1}
\paragraph{$B_1$}
Bubblesort rests on a key primitive: transposing or swapping
elements in an array.  The first system $B_1$ is designed
around this idea.  We assume $\id{swap}(a, i, j)$ denotes
the result of swapping $a_i$ with $a_j$ in $a$.  The state
vector of $B_1$ consists of an array $a$.  An action in
$B_1$ is of the form ${\sf swap}(i,j)$, where $0\leq i<
{j}<{n}$.  The dynamics of $B_1$ is captured by the
transition relation $a\ \xto{{\sf swap}(i,j)}{B_1}\ a'
\qquad \mbox{iff } a'=\id{swap}(a, i, j)$.
\newcommand{\swap}{{\sf swap}}

\paragraph{$B_1$ runs}
Here are example runs in $B_1$ starting with the array $[8,
  6, 7, 4]$.  The underlines identify elements being
swapped.

\begin{align}
[\underline{8}, 6, 7, \underline{4}]  \ \xto{\swap(0, 3)}{B_1}\ %
[4, 6, 7, 8]\label{run:b11}\\[1.5em] 
[8, \underline{6}, \underline{7}, 4]  \ \xto{\swap(1, 2)}{B_1}\ %
[\underline{8}, 7, 6, \underline{4}]  \ \xto{\swap(0, 3)}{B_1}\ %
[4, 7, 6, 8]\label{run:b12}
\end{align}

\begin{figure}
\caption{Online experiment based on transition system $B_1$
  showing the  swapping of elements \label{fig:swap}}
    \includegraphics[width=3.0in]{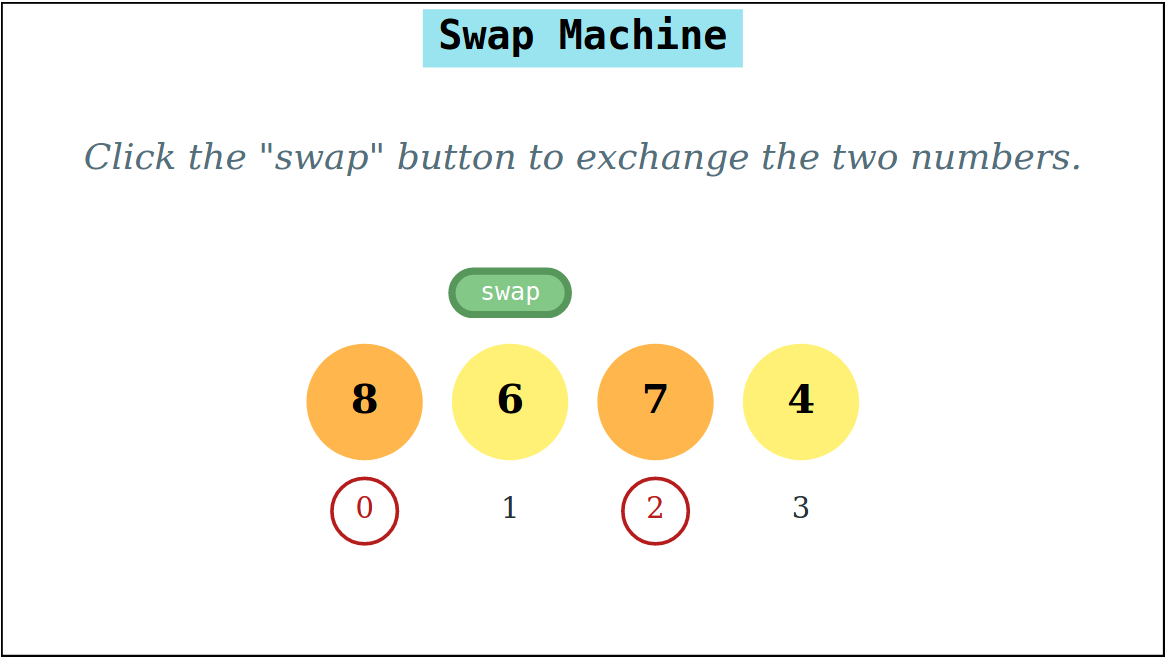}
\end{figure}

\paragraph{Virtual Lab experiment}
The runs may be easily created with paper and pencil by the
student and on the blackboard by the teacher.  In addition,
the student may create and explore these interactively
through a virtual experiment implementing this system.  (See
Figure~\ref{fig:swap}.)

\paragraph{Exploring $B_1$'s capability to sort}
Note that $B_1$ is interactive, deterministic and
non-terminating.  We are free to swap any two elements.
This can sometimes, result in a very short run to sort an
array (Run~\ref{run:b11}).  The freedom also makes $B_1$
versatile.  We can now use $B_1$ to not just sort, but also
reverse a sequence (Run~\ref{run:b12}).  With some practice,
the student could conjecture that $B_1$ may be used to
obtain {\em any} permutation of the array.  The more
inquisitive student may want to know why $B_1$ `works' for
sorting.  The more mathematically oriented student may be
encouraged to explore how this relates to the theory of
permutation groups\cite{2003-fraleigh-book}.


\subsubsection {Transition system $B_2$: ``Order Machine''}
\label{subsubsec:b2}
\paragraph{$B_2$ defn}
$B_1$ allows too much freedom and lacks direction.  Playing
with $B_1$, the student might realise that sorting can be
done by swapping only those elements that are out of order
(an operation we call `ordering').  The system $B_2$ captures
this insight.

\paragraph{$B_2$ actions}
The transition system $B_2$ has the array $a$ as its sole
state variable.  There is only one type of action in $B_2$:
${\sf order}(i, j)$, where, again $0\leq i < j < n$.  The
transition relation is $a\xto{{\sf order}(i,j)}{B_2} a'$ iff
$a_i > a_j\qquad \text{ and } a'=\id{swap}(a, i, j)$.


\paragraph{Conjecture correctness}
Interacting with the virtual experiment implementing $B_2$
should help the student conjecture that every run in $B_2$
terminates and the terminal state always is a sorted
array.  Of course, the mathematically inclined student may
wish to prove this property.  




\subsubsection{Transition System $B_3$: ``Order Adjacent machine''}
\label{subsubsec:B3}

\paragraph{$B_3$ definition}
How do we choose indices for ordering?  The simplest
strategy picks adjacent elements.  The transition system
$B_3$ incorporates this choice.  Therefore, $B_3$ has as its
state variable the array $a$ and single type of action ${\sf
  adj}(i)$ where $0\leq i < n-1$.  Its dynamics is described
by the transitions in $B_2$: $a \ \xto{{\sf
    adj}(i)}{B_3}\ a'$ iff $a \xto{{\sf order}(i, i+1)}{B_2}
a'$.


\subsubsection{Transition System $B_4$: ``Bubble'' Machine}
\label{subsubsec:B4}

\paragraph{$B_4$ definition}
$B_3$ still leaves undecided the strategy for selecting the
next action.  This problem is addressed by the system $B_4$,
which adopts a simple, linear traversal strategy to
automatically locate the next index.  Note that now the
choice of {\em which} index to consider for swapping
adjacent elements is no longer available.  It is automated
via an index variable $i$ in $B_4$.  $i$, initialised to $0$
is maintained as part of $B_4'$s state, along with the
sequence $a$.

\paragraph{${\sf inc}$ action in $B_4$}
To achieve the linear incremental strategy, $B_4$ comes
equipped with an action ${\sf inc}$.  ${\sf inc}$ orders the
adjacent elements at $i$ and $i+1$ and then automatically
increments $i$.  A sequence of ${\sf inc}$ actions sweep the
array $a$ starting from $i=0$.

\paragraph{${\sf reset}$}
A single sweep of the array is insufficient for sorting.
Therefore, the student is given the option of resetting the index
any time. A reset action heralds the beginning of a new sweep
cycle. This is accomplished by the ${\sf reset}$ action:

\paragraph{Transition relation}
For someone interested in observing only the array variable
in $B_4$, the view function $h_{B_4}$ is just
$h_{B_4}(a,i)=a$.  The transitions for $B_4$ are:
$(a, i) \xto{{\sf reset}}{B_4} (a, 0)$, 
$(a, i)\xto{{\sf inc}}{B_4} (a, i+1)$ iff  $i<n-1$  and
$a_i \leq a_{i+1}$, and $(a, i)\xto{{\sf inc}}{B_4} (a', i+1)$
iff $i<n-1$  and $a\xto{{\sf adj}(i)}{B_3}a'$.

\paragraph{Conjecturing the bubbling of the maximum element}
Playing with $B_4$ by tracing out a few runs, an observant
student may notice the following phenomenon:  the index $i$
carries along with it the maximum element in the array swept
so far.  The largest element `bubbles its' way towards the
right end of the array. 

\paragraph{$B_4$ terminiation}
However, the bubbling may be interrupted by a ${\sf reset}$
at any step.  Furthermore, the reset could be invoked
infinitely often.  $B_4$ is thus no longer a terminating
system.


\subsubsection{Transition System $B_5$: ``Bubblesort'' machine}
\label{subsubsec:b5}
A student could discover that repeated sweeps of the $B_4$
machine can be used to arrange the maximum element of an
array at position $n-1$, the next maximum at position $n-2$,
and so on.  This divides the array at any step into an
unsorted part followed by a sorted part.  Whenever the sweep
index $i$ reaches the boundary between the two parts, a
reset could be triggered.  The system $B_5$ implements this
strategy for resetting.  The state vector of $B_5$ now
includes an additional {\em boundary} variable $b$, that
ranges from $0$ to $n$.  A boundary value of $k$ means that
(a) all elements $a_k$ to $a_{n-1}$ are sorted and (b) $a_k$
is greater than or equal to all elements $a_0$ to $a_{k-1}$.

With this insight, it is now possible to define the dynamics
of $B_5$.  The state vector $(a, i, b)$ is initialised to
$(a^0, 0, n)$.  The sweep index $i$ varies from $0$ to
$b-1$.  At each step, the sweep index $i$ is incremented,
until it reaches $b-1$.  After that $i$ is reset to $0$.
Simultaneously, the boundary index $b$ is decremented.  The
system terminates when $b$ is less than or or equal to 1.
$B_5$ has only one action ${\sf next}$.  making it an
automated system.  The observable of $B_5$ is the array $a$.
The dynamics of $B_5$ is defined below:
\begin{align*}
  (a, i, b) & \xto{{\sf next}}{B_5}\ (a', i', b)  & %
   \text{iff } i < b-1\  \text{and } (a,i)\ \xto{{\sf inc}}{B_4}\ (a',i')\\
  (a, i, b) & \xto{{\sf next}}{B_5}(a', i', b-1) &%
  \text{iff }\ i = b-1 \text{ and}\ (a,i)\ \xto{{\sf reset}}{B_4}(a', i')
\end{align*}

\begin{figure}
  \caption{Screenshot of the Bubblesort algorithm (system
    $B_5$) displaying the values of array, index and
    boundary state variables during a sorting run.\label{fig:b5}}
  \includegraphics[width=3.0in]{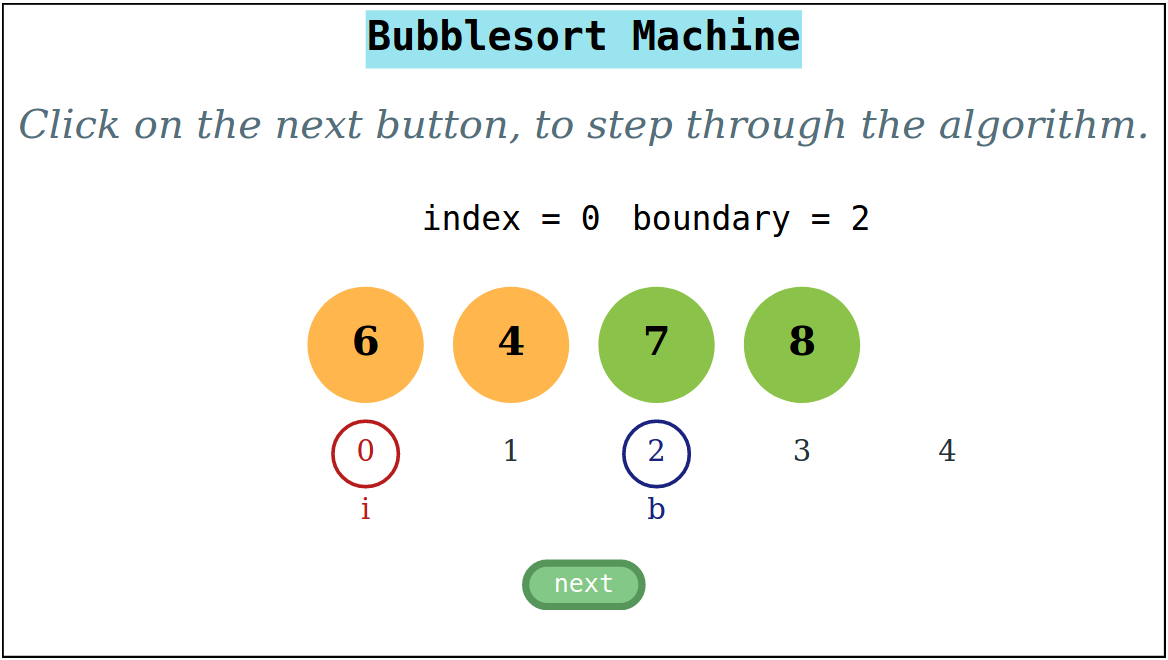}
  \end{figure}

An interactive experiment for $B_5$ allows the student to
step through the bubblesort algorithm.  A screenshot of the
algorithm in progress is shown in Figure~\ref{fig:b5}.

\paragraph{Termination of $B_5$}
Whenever $i$ reaches $b-1$, it is automatically reset, $b$
is decremented and as a result, the sorted segmented of the
array grows by one.  A formal correctness proof uses
induction on the transition systems is not difficult, but
omitted.

Table~\ref{tbl:runs} compares the runs amongst the five
transition systems $B_1$ to $B_5$ on the input array $[8, 7,
  6, 4]$.  It is now a simple exercise for the student to
improve the Bubblesort algorithm by incorporating a {\sf
  done?} boolean state variable that terminates the
algorithm as soon as there is a sweep with no swaps.






\begin{table}
\caption{The sorting of $[8, 6, 7, 4]$ witnessed by runs in
  the various transition systems $B_1$ to
  $B_5$.\label{tbl:runs}.  The boundary variable $b$ is part
  of only $B_5$'s state.  The index variable $i$
  (underlined) is part of only $B_5$'s and $B_4$'s state.}
\begin{tabular}{llllll}
\hline
\hline
State & B\(_{\text{5}}\) & B\(_{\text{4}}\) & B\(_{\text{3}}\) & B\(_{\text{2}}\) & B\(_{\text{1}}\)\\
\(a\), \(i\), \(b\) &  &  &  &  & \\
\hline
\([\underline{8}, 6, 7, 4], 0, 4\) & \({\sf next}\) & \({\sf inc}\) & \({\sf adj}(0)\) & \({\sf order}(0,1)\) & \({\sf swap}(0,1)\)\\
\([6, \underline{8}, 7, 4], 1, 4\) & \({\sf next}\) & \({\sf inc}\) & \({\sf adj}(1)\) & \({\sf order}(1,2)\) & \({\sf swap}(1,2)\)\\
\([6, 7, \underline{8}, 4], 2, 4\) & \({\sf next}\) & \({\sf inc}\) & \({\sf adj}(2)\) & \({\sf order}(2,3)\) & \({\sf swap}(2,3)\)\\
\([6, 7, 4, \underline{8}], 3, 4\) & \({\sf next}\) & \({\sf reset}\) &  &  & \\
\([\underline{6}, 7, 4, 8], 0, 3\) & \({\sf next}\) & \({\sf inc}\) &  &  & \\
\([6, \underline{7}, 4, 8], 1, 3\) & \({\sf next}\) & \({\sf inc}\) & \({\sf adj}(1)\) & \({\sf order}(1, 2)\) & \({\sf swap}(1,2)\)\\
\([6, 4, \underline{7}, 8], 2, 3\) & \({\sf next}\) & \({\sf reset}\) &  &  & \\
\([\underline{6}, 4, 7, 8], 0, 2\) & \({\sf next}\) & \({\sf inc}\) & \({\sf adj}(0)\) & \({\sf order}(0, 1)\) & \({\sf swap}(0,1)\)\\
\([4, \underline{6}, 7, 8], 1, 2\) & \({\sf next}\) & \({\sf reset}\) &  &  & \\
\([\underline{4}, 6, 7, 8], 0, 1\)\\
\hline
\end{tabular}
\end{table}

\subsection{Coding the transition systems into a program}
The process of translation consists of systematically coding
the actions of each transition system.  Each action maps to
a function whose arguments consist of the state variables.
The final transition system $B_5$ contains a single while
loop along with a termination condition.  A translation to
Python code is show below:
\label{subsec:bs-impl}
\lstset{language=Python,label= ,caption= ,captionpos=b,numbers=none, basicstyle=small}
\begin{lstlisting}[language=Python, basicstyle=\footnotesize]
# action from B_1            # action from B_2
def swap(a, i, j):           def order(a, i, j): 
 a[i], a[j] = a[j], a[i]      return swap(a, i, j)
 return a
# action from B_3            # action from B_4    
def adj(a, i):               def reset(a, i):
 return order(a, i, i+1)      return (a, 0)
# action from B_4            # action from B_5    
def inc(a, i):               def next(a, i, b):
 if (a[i] > a[i+1]):	      if (i < b-1):
  return (adj(a, i), i+1)      (a, i) = inc(a, i)
 else:	                       return (a, i, b)
  return (a, i+1)             else:
                               (a, i) = reset(a, i)
                               return (a, i, b-1)

def b5(a, i, b):             def bubblesort(a):
 while (b > 1):               (i, b) = (0, len(a))
  (a, i, b) = next(a, i, b)   (a, i, b) = b5(a, i, b)
    return (a, i, b)          return a
\end{lstlisting}

Prototype implementations of the five transitions systems
suggested in this paper are available
online\cite{2020-iticse-bubblesort-demo}.

\section{Conclusions and Future work}
\label{sec:conc}

We have introduced the algodynamics approach to teaching and
designing of algorithms.  We have illustrated the approach
using Bubblesort.  We have applied this approach on other
sorting and searching algorithms, algorithms on trees,
recursion and also concurrent algorithms.  These will be
reported elsewhere.  

For the student, the algodynamics approach shows a clear way
to think about the algorithm, in an interactive way.  It
encourages the teachers to construct their own pathways of
transition systems to explain the design of an algorithm.


Field level trials (with suitably designed teaching kits),
both for teachers and students are essential to test the
efficacy of this approach and remain to be done.
Incorporating the algodynamics approach in the syllabi of
the algorithms course will require the introduction of
transition systems earlier in the curriculum and integrating
it with the course on automata theory.  How this is to be
done remains to be explored.

\bibliographystyle{ACM-Reference-Format}
\bibliography{algorithm-papers,bubblesort-ref,other-refs,av-papers}


\end{document}